\pdfoutput=1

\documentclass[aps, prl, twocolumn]{revtex4-2}

\usepackage{amsmath}
\usepackage{amssymb}
\usepackage{color}
\def\mL{\mathcal{L}}
\usepackage{hyperref}

\def\lam{\lambda}
\def\flushboth{%
  \let\\\@normalcr   \@rightskip\z@skip \rightskip\@rightskip   \leftskip\z@skip
  \parindent 1.5em\relax}

\usepackage{verbatim}
\usepackage{graphicx}
\usepackage{subfig}
\newcommand{\bi}{\begin{itemize}}\newcommand{\ei}{\end{itemize}}
\newcommand{\be}{\begin{equation}}\newcommand{\ee}{\end{equation}}
\newcommand{\bml}{\begin{multline}}\newcommand{\emll}{\end{multline}}

\def\({\left(} \def\){\right)}\def\[{\left[} \def\]{\right]}
\newcommand{\bea}{\begin{eqnarray}}\newcommand{\eea}{\end{eqnarray}}

\def\ie{\begin{equation}\begin{aligned}}\def\fe{\end{aligned}\end{equation}}

\begin{document}

\title{On the spectra of strong turbulence on water surfaces}%\\[10pt]
\author{Gregory Falkovich}
\affiliation{Weizmann Institute of Science,  Rehovot 7610001 Israel}

\begin{abstract}
Here, I give a brief history of  the theory, with twists and turns, and list possible predictions for the spectra of the direct cascade of gravity waves on the water. The description of long waves is straightforward within the weak-turbulence theory developed by Zakharov and his school. As the energy cascade proceeds to shorter scales, the mean surface steepness grows, and wave-breaking events appear with increasing frequency. Considering surface cusps,  Phillips suggested a wavenumber spectrum (with a wrong magnitude as I argue here) and a frequency spectrum whose whole form is incorrect. The right form for the frequency spectral slope due to breaking waves was suggested by Kuznetsov. Here we estimate the magnitude and argue that the frequency spectrum is expected to continue unchanged through the whole cascade.  
The transition from weak to strong turbulence in the wavenumber and spatio-temporal spectra is  due to spectrally nonlocal interactions and happens earlier (at larger scales) than the dimensional estimate suggests. As a result, the magnitude of the strong-turbulence spectrum is not universal but depends on the pumping for a sufficiently long weak-turbulence cascade. The reason for nonlocality is likely that breaking of shorter waves is determined by the magnitude of the longest wave on whose crest they are running. 
\end{abstract}
\maketitle
{\bf Universal spectra hypothesis}. The subject of this note are the spectra of the squared surface elevation: $$\langle\eta^2\rangle=\int d^2k\langle|\eta_k|^2\rangle= \int \!\Phi(k,\omega)kdkd\omega=\int d\omega\,\Psi(\omega)\,.$$ 
We are interested in the asymptotic at large $k,\omega$ where we expect some degree of universality and scale invariance, so we look for power-law spectra, $\langle|\eta_k|^2\rangle\propto k^{-x}$ and $\Psi(\omega)\propto \omega^{-y}$. In such a limit, it is also tempting to assume spatio-temporal self-similarity: $\Phi(k,\omega)=\Psi(\omega)\omega^{-2a}f(\omega^a/k)=\langle|\eta_k|^2\rangle k^{-1/a}g(\omega^a/k)$ with
\be \omega\propto k^{1/a}\,,\quad a={y-1\over x-2}\,.\label{a}\ee

The first attempt to predict the spectra was straightforward and elegant: In 1958, Phillips suggested that ultimate short-wave asymptotic is universal, that is determined solely by the gravity $g$.  Simple count of dimensionalities  gives up to a numerical factor
\bea &\langle|\eta_k|^2\rangle\simeq k^{-4}\,,\label{P1}\\ &\Psi(\omega)\simeq g^2\omega^{-5}\ .\label{P2}\eea
The first formula seems physically transparent, as it corresponds to the Fourier transform of a wave cusp with an order-unity angle, giving $\eta_k\propto 1/k^2$. The wavenumber spectrum \eqref{P1} is so appealing since it is the first example of a critical-balance hypothesis: the mean squared surface slope (naive dimensionless nonlinearity parameter), $\epsilon_k=k^4\langle|\eta_k|^2\rangle$, is scale-independent and of order unity. On the contrary, the frequency spectrum \eqref{P2} had no physical picture behind it, but according to \eqref{a} gives $a=2$, which corresponds to the dispersion relation
$\omega=\sqrt{gk}$ valid for small-slope waves. Here we see the first contradiction, as the wavenumber spectrum assumes, on the contrary, order-unity slopes.

{\bf Weak turbulence}. The first consistent derivation of the spectra was made by Zakharov who developed the weak-turbulence theory for small-slope waves   \cite{ZLF,Naz}, where both the wavenumber and the frequency spectra can be obtained as exact solutions of the wave kinetic equation  \cite{Hass}:
 \bea &
{\partial n_k\over\partial t}= \!\! \int d{\bf p_2}d{\bf p_3}d{\bf p_4}\, \lam_{k234}^2\,n_k n_2n_3n_4\Big( \frac{1}{n_k} {+} \frac{1}{n_2}{-}\frac{1}{n_3} {-} \frac{1}{n_4} \Big)  
 \nonumber\\&\times
\delta({\bf k}{+}{\bf p_2}-{\bf p_3}{-}{\bf p_4}) \delta(\omega_{k} {+} \omega_{p_2}{-}\omega_{p_3}{-}\omega_{p_4})\,.\label{32}
\eea
Here we defined  $n_k=c_k\langle|\eta_k|^2\rangle$, where $c_k=\sqrt{g/k}$ is the phase velocity.  For surface water waves, the interaction coefficient $ \lam_{k234}$ has dimensionality $cm^{-3}$ that is proportional to the third power of the wavenumber (and does not depend on $g$); the full expression can be found in the Appendix. By a simple power counting, we see that the wavenumber spectrum carrying constant energy flux $P\sim \lambda^2n^3k^6$ must be $n_k\simeq P^{1/3}k^{-4}$. Zakharov and Filonenko substituded this into \eqref{32} and showed that all integrals there converge, and the right hand side turns indeed into zero \cite{ZF}. Since for weak turbulence $\Phi(k,\omega)\propto\delta(\omega-\sqrt{gk})$, then  
\bea & \langle|\eta_k|^2\rangle\simeq   ug^{-1/2} k^{-7/2}\,,\label{ZF1} \\& \Psi(\omega)\simeq ug\omega^{-4}\ .\label{ZF2} 
\eea
Here we followed tradition and introduced $u\equiv P^{1/3}$. Since $P$ is the energy dissipation rate per unit mass per unit area, then $u$ has the dimensionality of velocity (related to the speed of wind or wavemaker) and makes the subsequent formulas particularly transparent. Note that convergence of the integrals in \eqref{32} made (\ref{ZF1},\ref{ZF2}) local, that is independent of the pumping scale (yet they remain dependent on the pumping rate). The weak turbulence theory was expected to be valid for waves whose mean squared slope is less than unity or (equivalently) whose fluid acceleration $kv_k^2\simeq k(\eta_kk\omega_k)^2$ is less than $g$ for any $k$-wave. We estimate  from \eqref{ZF1}:  
\be\epsilon_k={kv_k^2\over g}\simeq {u\over c_k}=\sqrt{k\over k_n}\,,\qquad  k_n={g\over u^2}\,.\label{kn}\ee
We see that $\epsilon_k<1$ when $k<k_n$. The wavenumber $k_n$ is  when breaking of a single $k$-wave is expected.
%: the self-generated Doppler shift $kv_k$ is comparable to $\omega_k=\sqrt{gk}$ and the steepness is of order unity. 

The weak-turbulence spectrum \eqref{ZF1} matches the universal cusp spectrum \eqref{P1} at  $k_n$.
Similarly, the spectrum \eqref{ZF2} would go into \eqref{P2} at $\omega_n=\sqrt{gk_n}=g/u$. It is thus tempting to conclude that we have (\ref{ZF1},\ref{ZF2}) to the left and (\ref{P1},\ref{P2}) to the right of $k_n,\omega_n$. This is incorrect. The reason is the spectral nonlocality of the frequency spectrum and of multi-wave interactions.

{\bf Sweeping and $\omega$-spectrum}. First nonlocality is related to the frequency spectrum. The spectrum \eqref{P2} is self-contradictory, as stated above. 
Experiments and observations consistently show $\Psi(\omega)\propto\omega^{-4}$ (see e.g. \cite{PRZ}). Phillips in 1985 accepted the spectra (\ref{ZF1},\ref{ZF2}) (without ever mentioning Zakharov) and suggested that (\ref{P1},\ref{P2}) should be ``interred with dignity'' \cite{Phil85}. I would not do that for \eqref{P1}, since the cusps on a stormy sea can be seen by a naked eye. However, the data showed  $\omega^{-4}$ even when wave breakings were abundant. The simple observation which resolved this controversy was made by Kuznetsov in 2004, who noticed that cusps passing through any point produce jumps in time derivative of the elevation which give $\Psi(\omega)\propto \omega^{-4}$ \cite{K04}. Let us add to this that the magnitude of $\Psi(\omega)$ must be proportional to the mean squared speed of cusps running through a measurement point divided by the time between cusps, that is to   $\omega_kv_k^2$. This means that the frequency spectra cannot have a universal pumping-independent asymptotics. The main contribution into sweeping is due to longest waves (from the weakly-turbulent part of the spectrum), where $v_k^2\simeq  uc_k$ and  $\omega_kv_k^2\simeq ug$ is independent of $k$, so that  the frequency spectrum  \eqref{ZF1} must be the same for weak and strong turbulence. Sweeping is thus the first manifestation of spectral nonlocality making the frequency spectrum at however large $\omega$ dependent on the pumping rate at low $\omega$.

{\bf How straight are the cusps?} Could we now call it a day, assuming that the frequency spectrum is unchanged while the  wavenumber spectrum turns from the pumping-determined spectrum \eqref{ZF1} to the universal spectrum \eqref{P1} at $k_n$ where they match?  

Not yet. The Phillips wavenumber power law \eqref{P1} was questioned by Kuznetsov  \cite{K04}:  It may seem from the simple consideration above that it is enough to know the order of surface singularity to predict the spectrum: shock (jump) gives $\eta_k\propto 1/k$, cusp gives $\eta_k\propto 1/k^2$, etc; squaring it, one gets the wavenumber spectrum. This is true in one dimension only. At higher dimensions, the form of singularity matters. As was shown by Kadomtsev and Petviashvili for acoustic shocks \cite{KP}, Saffman for vorticity shocks \cite{S}, and generalized by Kuznetsov for general cases including surface cusps  \cite{K04}, the spectrum of randomly placed straight singularities must decay with $k$ faster by an extra factor $k^{1-d}$. Model computation illuminates this result. Consider a straight cusp of the length $L$ given by $\eta(x,y)=e^{-|x|/L-(y/L)^2}$ and denote $\theta$ the angle between the wavevector and the cusp normal. The Fourier tranform is given by
$|\eta_k|^2=[L^{-2}+(k\cos\theta)^2]^{-2}\exp[-(kL\sin\theta)^2]$. We see that each cusp line contributes to the spectrum a jet with an angular width decreasing as $1/kL$; averaging over the angle $\theta$, Kuznetsov obtained  at $kL\gg1$  \cite{K04}: 
\be  \langle|\eta_k|^2\rangle\simeq 1/ k^{5}L
%\,,\quad \Psi(omega)\simeq ug\omega^{-4}
\ .\label{Kuz} \ee 
If true{ \it and} if there is a self-similarity of $\Phi(k,\omega)$ then (\ref{a},\ref{ZF2},\ref{Kuz}) give $a=1$, that is $\omega\propto k$. Maxima of $\Phi(k,\omega)$ at straight lines in $(k,\omega)$-plane are frequently observed, yet it is unclear whether they dominate enough to provide such self-similarity at large $k,\omega$.
%Such spectrum would match \eqref{ZF1} and start  strong turbulence at the wavenumber $k_n^{1/3}L^{-2/3}$ which coincides with $k_n$ only if the transversal extend of the cusp is comparable to the wavelength of the breaking wave. 
Note, however, that \eqref{Kuz} is true only for straight fronts in real space, which is the case for acoustic shocks. I don't think that this is true for wedge-shaped breaking water waves, but it is ultimately for computations or measurements to settle. For non-straight cusps, I expect that the  Fourier spectrum is independent of the angle, as for a model of a  circular cusp,  $\eta({\bf r})=e^{-|{\bf r}-{\bf R}|/L}$, where $|\eta_k|^2=2\pi [L^{-2}+k^2]^{-2}$, giving the asymptotics  $k^{-4}$. This is similar to the enstrophy-cascade spectrum in 2d incompressible turbulence, which follows Kraichnan's $\langle |v_k|^2\rangle\propto k^{-4}$ rather than Saffman's $k^{-5}L$ despite the presence of the vorticity shocks, since the latter are not straight but curved \cite{KS}.

{\bf Interaction renormalization and nonlocality}. To establish the magnitude of the strong-turbulence wavenumber spectrum, one needs to know at which $k$ it matches the well-defined weak turbulence solution. That requires an honest calculation of the validity limit of weak-turbulence theory rather than the dimensional estimate \eqref{kn}. 
The first-order correction to the kinetic equation is the renormalization of the interaction vertex $\lambda_{1234}$. As was noticed by Rosenhaus and Falkovich \cite{RF}, even when the integrals in the kinetic equation converge for Zakharov solution, the next-order corrections could be magnified by divergences: nonlinearity is enhanced by nonlocality in non-equilibrium. The first (one-loop) correction to $\lambda_{1234}$ is obtained by the standard perturbation theory \cite{RS} and is as follows:
\bea\!\! \! \sum_{pq}\left[{  \lam_{12pq}\lam_{pq34}  (n_p\!+\!n_q ) \over  \omega_{1}\!+\!\omega_{2}\!-\!\omega_{q}\!-\!\omega_p\!+\!\imath\epsilon}+{ 4 \lam_{1p3q}\lam_{4q2p}  (n_q\!-\!n_p)
   \over \omega_{4}\!+\!\omega_{q}\!-\!\omega_{2}\!-\!\omega_p\!+\!\imath\epsilon}\right]\!\,.\label{bterm}
\eea
Substituting here the weak turbulence spectrum, $n_k\simeq uk^{-4}$, we obtain $\delta\lambda/\lambda\simeq ukk_p^{-1/2}=\epsilon_k(k/k_p)^{1/2}$.  We consider for simplicity the pumping (or spectral peak) wavenumber $k_p$ to be the smallest in the system and cut the divergence there.  It is important to note that the IR divergences are due to non-resonant interactions. This is why there are no divergences in the kinetic equation which containes only resonant interactions.

An important conclusion is that wavenumber spectrum must change from weak to strong turbulence when the relative correction is of order unity, that is at the wavenumber
\be k_*=g/uc_p=\sqrt{k_pk_n}\ .\label{star}\ee
Here $c_p=\sqrt{g/k_p}$ is the phase velocity of the longest (and fastest) waves. 
The physics behind that second nonlocality could be that wave breaking starts not for isolated waves with the critical slope (and wavenumber $k_n$) but for waves running on the crests of long waves  \cite{IV} and  at a level of steepness significantly smaller than  
critical \cite{break1}. Indeed, the fluid flow of a large wave distorted by a small wave gives the fluid acceleration $kv_p^2=kuc_p$, which is equal to $g$ for $k=k_*$. At the same wavenumber, the Doppler shift due to a large wave, $kv_p=k\sqrt{uc_p}$, exceeds the linear frequency $\sqrt{gk}$. That spectral nonlocality may also shed some light on the seemingly absurd relation between the statistical distribution of breaking waves over speeds and the weak-turbulence spectrum \eqref{ZF1} where no breaking occurs  \cite{Phil85}. Direct computations seems to support this relation \cite{Deike23,Deike25}. That could be explained if indeed breaking waves preferably ride on longer waves from the weakly-turbulent part.  

{\bf Frequency renormalization}: $\tilde\omega_k=\sqrt{gk}+2\Sigma_k$. The expansion in powers of nonlinearity is as follows:
\bea\!\!\!\!\!\!\Sigma_k=\sum_{\bf q}\! \lambda_{kqkq}n_q%\\& 
- \sum_{\bf pqr}{ \lam_{kpqr}^2 (n_pn_q+2n_qn_r )
   \over  \omega_{k}\!+\!\omega_{p}\!-\!\omega_{q}\!-\!\omega_{r}+\imath\epsilon}\!+\!\ldots\label{Sigma}
\eea 
% Up to the second order, we are not interested in the imaginary part (dissipation) bacuse it is already included into the kinetic equation.
The first term $\Sigma_1$ in \eqref{Sigma} is degenerate since $ \lambda_{\bf kqkq}=q({\bf k}\cdot{\bf q})$ for $q\ll k$ as can be found from the below expression for $\lambda_{ \textbf{pqkr}}$ \cite{K94,PRZ}. The IR log-divergent contribution,
$\Sigma_1=2\lambda  ({\bf k\cdot R})$, is proportional to the vector ${\bf R}= \int {\bf  q}qn_qd{\bf q}$. For a momentum-carrying distribution, the renormalization is nonzero:  $ \Sigma_1\simeq uk\cos\theta\log(k/k_p)$. That, however, does not change the renormalized kinetic equation, since the linear in ${\bf k}$ terms cancel out both in the frequency delta function and in the denominators of the loop integrals. For isotropic spectra, $\Sigma_1=7\pi^2 uk/8$, that is positive and linear in $k$ \cite{Vlad}. Absence of a power-law divergence, means that $\Sigma_1\simeq \omega_k$ at $k\simeq k_n$. That frequency renormalization does not directly influence the wavenumber spectrum of the energy cascade yet could be probably observed in $\Psi(k,\omega)$.   %Note that experiments also show maximum of $\Psi(k,\omega)$ at $\omega=kc_p/2$, which corresponds to a sweeping with the group velocity of the spectral peak, see e.g. \cite{Shavit25}.

The second term gives $\Sigma_2\simeq -u^2k^{5/2}/k_p\sqrt{g}$  - because of power IR divergences, the contribution into the integral of \eqref{ZF1} exceeds that of\eqref{Fal1}. We find that $\Sigma_2\simeq \omega_k$ at the same $k_*$ when the vertex renormalization starts to matter. We conclude that weak turbulence must turn into strong at $k_*$. 

{\bf The new spectra}. The strong-turbulence spectrum $\langle|\eta_k|^2\rangle=Ak^{-4}$ that match \eqref{ZF1} at $k_*$ is as follows:
\bea  &\langle|\eta_k|^2\rangle\simeq k^{-4}\sqrt{u\over c_p}%=k^{-4}{v_p\over c_p}
=k^{-4}\sqrt{k_p\over k_*}=k^{-4}\epsilon_p^{1/2}
\ .\label{Fal1}\\  &\Psi(\omega)\simeq ug\omega^{-4}\,\label{Fal2} \eea
Would one had chosen \eqref{Kuz},  $\langle|\eta_k|^2\rangle=Bk^{-5}$, then $B\propto u^{-1/2}$. This is a very dramatic statement  - the decrease of the spectrum magnitude with increasing pumping because of an extreme nonlocality. While  seems paradoxical, such behavior is not impossible - it was recently predicted (bit not yet observed) for spin-wave turbulence \cite{spin1}.
 Note that the frequency spectrum is proportional to $u=P^{1/3}$, while the wavenumber spectrum --- to $P^{1/6}$.  The magnitude of \eqref{Fal1} is suppressed relative to \eqref{P1} since the breaking waves occupy less area: $k_p/k_*$ rather than $k_p/k_n$. Apparently, \eqref{Fal1} is valid only for $\epsilon_p=k_p/k_*\leq1$. When $k_*$ reaches $k_p$ so that turbulence is strong already at the pumping scale and breaking waves are space-filling, the universal spectrum \eqref{P1} must be realized on the whole interval $k\geq k_p$. At higher pumping, all the excess of energy spills into foam, and the
turbulence level saturates:
\bea &\langle|\eta_k|^2\rangle\simeq k^{-4}\,,\label{Fal3}\\ &\Psi(\omega)\simeq c_pg\omega^{-4}\ .\label{Fal4}\eea
The  wavenumber (Phillips) spectrum is truly universal, while the frequency spectrum does not depend on pumping but depends on the speed of the fastest waves. It is unclear to me whether the frequency spectrum continues to be $\Psi(\omega)\simeq ug\omega^{-4}$ when a wavemaker with the frequency $\omega_p$ generates flow with the surface speed $u$ exceeding $c_p$.

One needs numerics to establish how long spectral interval must be for the nonlocal interactions to  dominate. Preliminary data by Vladimirova show that, depending on the configurations of wavevectors, local and nonlocal interactions could give corrections of opposite signs, and that  one needs at least two decades for nonlocal effects to dominate $\delta\lambda_{1234}$ for comparable wavenumbers  \cite{Vlad}. One needs to compute numerically the overall contribution to $dn_1/dt$ of the one-loop correction.

{\bf Symmetries of the different spectra.} Original dynamical equations are invariant with respect to the transformation $\eta_k(t)\to \lambda^{2}\eta_{\lambda k}(\lambda^{-1/2}t)$ which corresponds to 
\be\Psi( \lambda k, \lambda^{1/2}\omega)=\lambda^{-5/2}\Psi(k,\omega) \,.\label{sym1}\ee
The hypothetical spectra (\ref{P1},\ref{P2}) respect this symmetry of the original equations \cite{NZ}.
Pumping imposes the scale $k_p^{-1}$ and the time $u/g$. The spectra (\ref{ZF1},\ref{ZF2}) are local, that is independent of $k_p$, yet they break the symmetry \eqref{sym1}), which can be restored by rescaling the flux $u\to \lambda^{-1/2}u$. The wavenumber spectrum \eqref{Fal1} depends on both the pumping magnitude and the scale.  The ultimate wavenumber spectrum (\ref{P1},\ref{Fal3}) is universal, while the frequency spectrum depends on $k_p$.

Are there self-similarities in $k,\omega$-plane? The weak-turbulence spectrum is concentrated on a single line in $(k,\omega)$-plane: $\Phi(k,\omega)=ug^{-1/2} k^{-7/2}\delta\left(\omega-\sqrt{gk}\right)$, that is self-similar with $a=2$. Strong-turbulence spectrum is unlikely to concentrate on a single line and even have any space-time self-similarity. Indeed, applying \eqref{a} to (\ref{Fal1},\ref{Fal2}) suggest not  $\omega=\sqrt{gk}$ but  $\omega\sim k^{2/3}\varepsilon^{1/3}$, where $\varepsilon(u)=  g(uc_p)^{1/2}$. Such Richardson self-similarity (squared distance proportional to the cube of time) is unlikely to be observed since it contradicts the frequency renormalization described by (\ref{Sigma}).  This renormalization is generally complex, that is describes spectral broadening as well, most likely as $\Sigma_k\propto k^{3/2}$. In addition, sweeping by longer waves must produce $\omega\propto k$. The spectrum $\Phi(k,\omega)$ then is unlikely to be self-similar in the strong-turbulence regime.

Let us briefly discuss the spectrum anisotropy. Within weak-turbulence theory, the isotropic spectrum is stable with respect to small angular perturbations \cite{ZLF}. The slowest-decreasing is the first angular harmonic proportional to the wave momentum flux: the $\cos\theta$-correction to \eqref{ZF1} decays as $k^{-4}$ and to \eqref{ZF2} as $\omega^{-5}$ \cite{KK}. The latter should not be mistaken for \eqref{P2}. This isotropization at small scales is supported by the  numerics of the kinetic equation  (see e.g. \cite{BGP}). On the other hand, observational data on wind-driven waves show bimodal angular structure developing at higher $k$ \cite{Hwa}. That could be that strong-turbulence spectra (of breaking waves) tend to break into narrow jets. For review of experiments, see \cite{Rev}.

The nonlocalities, derived above, may shed light on the remarkable observation by Korotkevich that long-wave condensate plays a role in formation of strong-turbulence spectrem $k^{-4}$ \cite{Kor08}. Indeed, our derivations show that both $\delta\lambda_{1234}$ and $\Sigma_k$ diverge for spectra steeper than $n_k\propto k^{-7/2}$ which is true for all spectra considered including the inverse cascade $n_k\propto k^{-23/6}$. 

{\bf Conclusion}. 

The most important and readily verifyable prediction is that the amplitude of the Phillips spectrum is non-universal but grows with the pumping strength until it saturates.

Formulas (\ref{star},\ref{Fal1},\ref{Fal2},\ref{Fal4}) are the main claims of this note. They are plausible suggestions based on the computation of one-loop corrections and on physical arguments. At this point it is unclear if the spectra  (\ref{Fal1},\ref{Fal2},\ref{Fal4}) are realized and whether they correspond to a critical balance and a constant energy flux.
\vskip 0.2cm

The predictions possibly amenable for experimental and numerical verification:

1. The deviations from weak turbulence start at $k_*$ rather than $k_n$. The deviations could be observed from either steepining of $\langle|\eta_k|^2\rangle$ or straightening and broadening of the maxima of $\Phi(k,\omega)$ in $k,\omega$-plane. 

2. The frequency spectrum $\Psi(\omega)$ is not expected to change upon transition to strong turbulence, yet the wavenumber-frequency spectrum $\Phi(k,\omega)$ must change significantly. One possible form of spectral broadening is given by \eqref{Sigma} where we now substitute \eqref{Fal1}:
\be {\rm Im}\,\Sigma_k\propto \sqrt{gk{u\over c_p}}{k\over k_p}\,\label{Sigma2}\ee

3. The wavenumber spectrum of strong turbulence, \eqref{Fal1},  has the Phillips scaling yet non-universal amplitude dependent both on the pumping rate and the peak wavenumber. An alternative prediction is \eqref{Kuz} where one needs to put either $L\simeq k_n^{-1}\propto u^2/g$ or $L\simeq k_*^{-1}=uc_p/g$ (for a long weak-turbulence cascade). In both cases, the amplitude of such spectrum is expected to {\it decrease} with increasing pumping $u$.  The simplest way to distinguish between the alternatives is to see whether the strong-turbulence spectrum increases or dicreases with increasing pumping. 
\vskip 0.2cm

From the analytic theory perspective, the following questions naturally appear:

1. If indeed the wavenumber spectrum decays faster with $k$ when the nonlinearity grows, that must mean that the effective interaction contributing the energy cascade is getting stronger. The relative vertex renormalization $\mL=\delta\lambda_{1234}/\lambda_{1234}$ where $\delta\lambda_{1234}$ is given by \eqref{bterm}, must be computable by combining analytics and numerics.  Since different confugurations of wavevectors give different signs (for instance, it seem that collinear give positive $\mL$  that is enhancement\cite{Vlad}), one needs to integrate over 2,3,4 to get the sign of the overall contribution to $dn_1/dt$ and see how long must be the spectral interval for nonlocal contribution to dominate.

2. Is there an equation of which either $\langle|\eta_k|^2\rangle\propto k^{-4}$ or $k^{-5}$ is a solution?
Dyachenko and Newell argued that $k^{-4}$ does not correspond to a constant flux but rather is a balance between wave interaction and dissipation \cite{DN}. I believe to the contrary, inspired by the recent success in deriving the critical-balance spectra for focusing Nonlinear Schr\"odinger equation \cite{VRF1} and spin waves \cite{spin1} as constant-flux solutions of renormalized kinetic equations. 

3. Critical balance is where the dimensionless degree of nonlinearity is independent of $k$. Remind that this degree is not the naive $\epsilon_k$ but $\mL$. For any $n_q=Aq^{-\gamma}$ with any meaningful $\gamma$ ($\gamma>7/2$), we have from \eqref{bterm}:  $\mL\simeq k^{3/2}Ak_p^{7/2-\gamma}/\tilde\omega_k$. That could be independent of $k$ and $u$ if one substitute $\tilde\omega_k$ in the denominator by $\Sigma_k\propto k^{3/2}$, see \eqref{Sigma2}. Could it be justified by any plausible summation? For example, bubbles for the vertex renormalization with $\omega_k\rightarrow\Sigma_k$ inside every bubble.

Assuming  spectrum \eqref{Fal1}, could one analyze the whole perturbation series and establish leading diagrams in the renormalization of the vertex and  the mass operator $\Sigma$?  Could one then self-consistently derive a close kinetic equation, whose solution is \eqref{Fal1}? Does this solution correspond to a constant energy flux?

4. All the infrared divergences show that the multi-mode interaction of short nonlinear waves is mainly with the long weakly nonlinear waves whose statistics can be assumed Gaussian. This suggests an effective renormalization scheme based on the explicit averaging over the long waves from the outset.
\vskip 0.2cm
%I thank Vladimir Rosenhaus, Natalia Vladimirova, Eugene Kuznetsov, Luc Deike and Miguel Onorato for useful discussions. The work  was supported by the Excellence Center at WIS and by the grants of BSF and ISF. 
I apologize for many omissions in the unsystematic list of refences below.

%\bibliographystyle{utphys}
%\bibliography{ConvergenceTurbbib}

\vskip 0.1cm
\vskip 1 cm \appendix{Appendix}

{\bf The interaction  vertex} is a function $\lambda_{\textbf{pqkr}}$ of four two-dimensional vectors with a zero sum: ${ \textbf{p}}=(a,b)$, $\textbf{q}=(c,d)$, ${ \textbf{r}}=(r,0)$, $\textbf{k}=\textbf{p}+\textbf{q}-\textbf{r}=(a+c-r, b+d)$. 

Let us define $Q_{\textbf{kpq}}=(q/ kp)^{1/4}\!(kp+  \textbf{k}\cdot \textbf{p}  )$, then 
\bea&U_{\textbf{kpq}}=-Q_{\textbf{-k,p,q}}-Q_{\textbf{-k,q,p}}+Q_{\textbf{p,q,-k}}\,,\nonumber\\
&\!\!U^\prime_{\textbf{kpq}}=Q_{\textbf{kpq}}+Q_{\textbf{kqp}}+Q_{\textbf{pqk}},\nonumber\\&  A_{\textbf{kpq}}= -{ U_{\textbf{kpq}}\over \sqrt{k}-\sqrt{p}-\sqrt{q}},\ A^\prime_{\textbf{kpq}}= -{ U^\prime_{\textbf{kpq}}\over \sqrt{k}+\sqrt{p}+\sqrt{q}}\label{QU}\nonumber\\ 
 &\!\!Z_{0123}=U^\prime_{01-0-1}A^\prime_{23-2-3}+ U_{0+101}A_{2+323}-U_{020-2} A_{313-1}\nonumber\\ &\!\!\!\!-U_{202-0} A_{131-3} -U_{030-3} A_{212-1}-
U_{303-0}A_{121-2}\,.\label{la2}\nonumber
\\&V_{0,1,2,3}=\left({k_2k_3\over k_0k_1}\right)^{1/4}\bigl[2k_0^2k_1+2k_1^2k_0-k_0k_1(| \boldsymbol{k_0+k_2} |\nonumber\\ &+| \boldsymbol{k_1+k_2} |+| \boldsymbol{k_0+k_3} |+| \boldsymbol{k_1+k_3} |\bigr]\,.
\label{V}\nonumber \\ &\!\!W_{\textbf{pqkr}}=V_{\textbf{-p,-q,k,r}}+ V_{\textbf{k,r,-p,-q}} - V_{\textbf{-p,k,-q,r}} \nonumber\\&- V_{\textbf{-q,k,- p,r}} - V_{\textbf{-p,r,-q,k}} - V_{\textbf{-q,r,-p,k}}\,,\label{W}\nonumber\\&\!\!\!\!\!\lambda_{ \textbf{pqkr}}=W_{\textbf{pqkr}}\!+\!{1\over4}\left( Z_{\textbf{pqkr}}\!+Z_{\textbf{qpkr}}\!+Z_{\textbf{krpq}}\! +Z_{\textbf{rkpq}}\right)\,\label{la1}\nonumber
\eea\vskip 1cm

{\bf Capillarity} contributes the frequency $\omega_k^2=gk [1+(k/k_c)^2]$ where $k_c^2=q/\sigma$. Note that dimensional nonlinearity parameter $k^4\eta_k^2$ depends neither on $g$ nor on $\sigma$. It is unclear how capillarity modifies the wavenumber spectrum $\eta_k\sim k^{-2}$. If $k_c<k_*$, then capillarity takes over from gravity in the weak-turbulence regime. Cubic nonlinearity $\eta v^2$ gives the three-wave interaction coefficient $V_{k12}\simeq  \omega_k^{1/2}k^{3/2}$. Together with the decay dispersion law $\omega_k=(\sigma k^3)^{1/2}$, they give the spectrum $n_k=u^{3/2}k^{-2}/V_k$, which corresponds to $\eta_k^2\simeq k^{-4} (u/c_k)^{3/2}=
u^{3/2}\sigma^{-1/4}k^{-17/4}$. Cubic nonlinearity parameter is $\eta v^2/\sigma (\nabla\eta)^2$. Only if we relate velocity and elevation by the linear relation, $v=\omega\eta$, that coincides with the internal nonlinearity parameter $\nabla\eta$.  Naively estimated, such nonlinearity $\epsilon_k=k^4\eta_k^2=(u/c_k)^{3/2}\propto k^{-3/4}$ decays with $k$, which suggests that such turbulence is weak at however large $k$. To check if it is true, one needs to compute the corrections to the frequency and the vertex of interaction; they behave according to the estimate if the respective integrals converge.

It is straightforward to show that the integral for the frequency renormalization,  $$\Sigma_k=\int |V_{k1k-1}|^2n_1(k^{3/2}-|k-k_1|^{3/2}-k_1^{3/2}+\imath\varepsilon)^{-1}k_1dk_1,$$ converges at $k_1\to0$, when  $ |V_{k1k-1}|^2\propto k_1^{m_1}k^{2m-m_1}= k_1^{7/2}k$. The imaginary part of $\Sigma$ give the $k-\omega$ spectrum broadening, which indeed was found numerically to scale according to the dimensional expectation: $\Sigma_k\propto k^{3/4}$ \cite{Cap}.  The one-loop corrections to the kinetic equation were computed in
\cite{loop}  for the 3-wave case. For $\lim_{q\to0}V_{kq,k+q}\propto q^{m_1}k^{2m-m_1}$, one finds the following IR properties:
For $V_a,V_b$ at $p_5\to0$, the first lines give $k_0^{m_1+d-\gamma}=k_0^{m_1-m}$. The same convergence one finds in the second and third lines because the two $n$-factors and two $\omega$-denominators are each proportional to $p_5$ and cancel each other. 
The propagator terms don't have cancellations in the $n$-terms, but have one-power cancellations in the last terms of $P_a+P_b$ and $P_c+P_d$, which makes $P_a+P_b+P_c+P_d\propto k_0^{m_1-m-1}$. All converge for capillary waves. We conclude that if the pumping excites weak capillary turbulence, it is getting weaker along the cascade. What if pumping excites strong turbulence?
%One needs to check if the one-loop correction is enhanced by any IR divergence.

Without gravity, capillarity allows an ultimate Crapper wave, when the surface self-touches enclosing of air bubbles in the trough \cite{crap}. That corresponds to a shape singularity $\eta\propto \sqrt{|x|}$, where the derivative turns into infinity at $x=0$. The capillary singularity is stronger than the cusp, and suggests slower-decaying Fourier spectrum: $\eta_k^2\propto k^{-3}$. It is likely that on such a wave, the radial acceleration $v^2/\eta$ is comparable with the capillarity acceleration $\sigma k^2$, so that the nonlinearity parameter  $\eta v^2/\sigma (\nabla\eta)^2\simeq 1$.

Such capillary waves of sufficiently high (yet sub-ultimate) amplitude have negative-slope parts, which are impossible for a finite gravity. Therefore, gravity-capillary waves are always smooth --- gravity prevents bubbles while capillarity rounds cusps making their radius of curvature $1/k_c$ so that capillary acceleration is equal to $g$. How exactly capillarity modifies Phillips spectrum at $k>k_c$?

%Note that $\lambda$ changes sign depending on the orientation and ratios.
\end{document}